\documentclass[aps,twocolumn,preprintnumbers,amsmath,amssymb,superscriptaddress,floatfix,nofootinbib]{revtex4}
\usepackage{graphicx,color,dcolumn,booktabs,bm}
\usepackage{longtable,lscape}
\usepackage{txfonts}
\usepackage{overpic}
\usepackage{epsfig}
\usepackage{amssymb}
\usepackage{rotating}
\usepackage{epstopdf}
\usepackage{indentfirst}
\usepackage{feynmf}   
\usepackage{slashed}  
\usepackage{cases}
\usepackage{color}
\usepackage{multirow}
\usepackage{graphicx,color,dcolumn,booktabs,bm}
\usepackage{cases}
\usepackage{array}

\graphicspath{{Figures/}} %

\usepackage[colorlinks, citecolor=blue,anchorcolor=red,menucolor=red, linkcolor=red,filecolor=red,runcolor=red,urlcolor=blue,frenchlinks=red]{hyperref}

\begin{document}

\title{Masses of doubly heavy tetraquarks $T_{QQ^\prime}$ in a relativized quark model}

\author{Qi-Fang L\"u} \email{lvqifang@hunnu.edu.cn}
\affiliation{  Department
of Physics, Hunan Normal University,  Changsha 410081, China }

\affiliation{ Synergetic Innovation
Center for Quantum Effects and Applications (SICQEA), Changsha 410081,China}

\affiliation{  Key Laboratory of
Low-Dimensional Quantum Structures and Quantum Control of Ministry
of Education, Changsha 410081, China}
\author{Dian-Yong Chen} \email{chendy@seu.edu.cn}
\affiliation{School of Physics, Southeast University, Nanjing 210094, China }
\author{Yu-Bing Dong} \email{dongyb@ihep.ac.cn}
\affiliation{Institute of High Energy Physics, Chinese Academy of Sciences, Beijing 100049, China}
\affiliation{Theoretical Physics Center for Science Facilities (TPCSF), CAS, Beijing 100049, China}
\affiliation{School of Physical Sciences, University of Chinese Academy of Sciences, Beijing 101408, China}

\begin{abstract}

In the present work, the mass spectra of doubly heavy tetraquarks $T_{QQ^\prime}$ are systematically investigated in a
relativized quark model. The four-body systems including the Coulomb potential, confining potential, spin-spin interactions, and
relativistic corrections are solved within the variational method. Our results suggest that the $IJ^P=01^+$ $bb \bar u \bar d$ state
is 54 MeV below the relevant $\bar B \bar B$ and $\bar B \bar B^*$ thresholds, which indicates that both strong and electromagnetic decays are forbidden, and thus this state can be a stable one. Its large hidden color component and small root mean square radius demonstrate that it is a compact tetraquark rather than a
loosely bound molecule or point-like diquark-antidiquark structure. Our predictions of the doubly heavy tetraquarks may provide valuable
information for future experimental searches.

\end{abstract}

\maketitle

\section{Introduction}{\label{introduction}}

In the past two decades, plenty of new resonances have been observed in the hadronic physics, and some of them can be hardly classified
into the conventional hadron sectors, i.e., mesons and baryons~\cite{Tanabashi:2018oca}. These exotic structures have attracted extensive theoretical and experimental
interests due to their enigmatic properties~\cite{Chen:2016qju,Hosaka:2016pey,Richard:2016eis,Lebed:2016hpi,Ali:2017jda,Esposito:2016noz,
Guo:2017jvc,Olsen:2017bmm,Karliner:2017qhf,Liu:2019zoy,Brambilla:2019esw}. To describe their inner structures, new effective degree of
freedom are introduced to go beyond the traditional quark-antiquark and three-quark configurations. The experimental observations of
charged quarkonium-like states $Z_{c(b)}$~\cite{Choi:2007wga,Aaij:2014jqa,Belle:2011aa,Ablikim:2013mio,Liu:2013dau} and pentaquark states
$P_c$~\cite{Aaij:2015tga,Aaij:2019vzc} provide strong evidences for the existence of the exotic hadrons in QCD. Besides these hidden
charm and bottom ones, it is also expected that the open flavor exotic states should exist. However, the experimental searches
for these flavored exotic hadrons were beset with difficulties and obstacles, and the experiences of failures, such as
$\Theta^+(1540)$~\cite{Nakano:2003qx} and $X(5568)$~\cite{D0:2016mwd,Aaij:2016iev}, have casted a shadow over this research area.

The situation began to change in 2017, when a doubly heavy baryon $\Xi_{cc}^{++}$ was observed by the LHCb
Collaboration~\cite{Aaij:2017ueg}.  Although the $\Xi_{cc}^{++}$ is regarded as a $S-$wave conventional baryon, it provides an excellent
opportunity to examine the interactions between two heavy quarks and search for more doubly heavy quark systems. Indeed, based on
the mass of $\Xi_{cc}^{++}$, the mass spectra of doubly heavy tetraquark states $T_{QQ^\prime}$ were studied
subsequently, which indicate that there should exist at least one stable flavored exotic tetraquark $bb \bar u \bar
d$~\cite{Karliner:2017qjm,Eichten:2017ffp}.

Actually, the doubly heavy tetraquarks $T_{QQ^\prime}$ have been discussed for a long time. Before the observation of $\Xi_{cc}^{++}$,
there have been a number of theoretical works on the doubly heavy tetraquarks. Various approaches, involving quark
models~\cite{Karliner:2017qjm,Eichten:2017ffp,Ballot:1983iv,Lipkin:1986dw,Vijande:2009ac,Luo:2017eub,Ebert:2007rn,Semay:1994ht,
Zouzou:1986qh,Heller:1986bt,Carlson:1987hh,SilvestreBrac:1993ss,SilvestreBrac:1993ry,Pepin:1996id,Brink:1998as,Vijande:2003ki,
Zhang:2007mu,Vijande:2009kj,Yang:2009zzp}, QCD sum rules~\cite{Navarra:2007yw,Dias:2011mi,Chen:2013aba,Du:2012wp}, and lattice
QCD~\cite{Ikeda:2013vwa,Bicudo:2015kna,Francis:2016hui,Bicudo:2017szl}, were adopted to estimate their mass spectra. Due to the lack of
experimental information on the doubly heavy systems, it is difficult to distinguish those numerous results. Also,
several works have investigated their production mechanism, which should be helpful for experimental searches
~\cite{Gelman:2002wf,Janc:2004qn,DelFabbro:2004ta,Yuqi:2011gm,Hyodo:2012pm}. Lately, stimulated by the observation of $\Xi_{cc}^{++}$,
the studies on doubly heavy systems were revived and have interested plenty of theorists and experimenters. In particular, the
properties for the doubly heavy tetraquarks, such as their masses, decays, and production rates, have been extensively discussed
in the past years~\cite{Mehen:2017nrh,Wang:2017dtg,Yan:2018gik,Ali:2018ifm,Xing:2018bqt,Agaev:2018vag,
Ali:2018xfq,Park:2018wjk,Agaev:2018khe,Francis:2018jyb,Junnarkar:2018twb,Deng:2018kly,Caramees:2018oue,Agaev:2019qqn,
Sundu:2019feu,Maiani:2019cwl,Zhu:2019iwm,Maiani:2019lpu,Fontoura:2019opw,Agaev:2019kkz,Leskovec:2019ioa,Liu:2019yye,
Hernandez:2019eox,Yang:2019itm,Tang:2019nwv,Agaev:2020dba,Wang:2020jgb,Tan:2020ldi}. Within different frameworks, these studies present
distinctive results and conclusions. However, almost all the works agree that the isoscalar $T_{bb}$ state should be stable against its
strong and electromagnetic decays. The binding energy relative to the $\bar B \bar B^*$ threshold is predicted to be more than 100 MeV
by most studies, which is deeply bound and leads to a compact configuration.

Within the framework of quark models, the previous studies were mainly based on the nonrelativistic quark potential models or simple
quark models. Since the doubly heavy tetraquarks also include two light antiquarks, the relativistic corrections for the mass spectra
may be significant. For instance, the masses of doubly heavy tetraquarks are calculated
within a relativistic quark model under the diquark approximation~\cite{Ebert:2007rn}. However, the four-body calculations together with relativistic
effects have not been done in the literature. Therefore, before making a final conclusion on the isoscalar $T_{bb}$ state, it is
essential to perform a calculation in  a relativized quark model with few-body method for the doubly heavy tetraquark spectra.

In this issue, we investigate the mass spectra of doubly heavy tetraquarks $T_{QQ^\prime}$ in the relativized quark model proposed by
Godfrey, Capstick, and Isgur~\cite{Godfrey:1985xj,Capstick:1986bm}. This model has been extensively adopted to study the properties of
conventional hadrons and it may give a unified description of different flavor sectors. Also, under the diquark approximation, the
authors have employed the relativized quark model to deal with the tetraquark states and achieved satisfactory results~\cite{Lu:2016cwr,
Lu:2016zhe,Lu:2019ira,Anwar:2017toa,Anwar:2018sol,Bedolla:2019zwg,Ferretti:2020ewe}. Thus, the relativized quark model is suitable for us
to deal with the doubly heavy tetraquarks, where all the heavy-heavy, heavy-light, and light-light quark interactions are involved.
For the first time, we extend the relativized quark model to investigate the double heavy tetraquark spectra by solving a four-body Schr\"{o}dinger-type equation. With the present extension, the tetraquark, as well as the conventional hadrons  can be described in a uniform frame.

This paper is organized as follows. The framework of relativized quark model and few-body method are introduced in Sec.~\ref{model}. The
results and discussions of doubly heavy tetraquark spectra are given in Sec.~\ref{results}. A summary is presented in the last section.

\section{Model}{\label{model}}

\subsection{Hamiltonian}

To calculate the mass spectra of doubly heavy tetraquarks $T_{QQ^\prime} \equiv Q Q^\prime \bar q \bar q^\prime$, the relativized
Hamiltonian should be constructed. Similar to the procedures of the conventional mesons and baryons~\cite{Godfrey:1985xj,Capstick:1986bm},
the relativized Hamiltonian for a $Q Q^\prime \bar q \bar q^\prime$ tetraquark state can be written as
\begin{equation}
H = H_0+\sum_{i<j}V_{ij}^{\rm oge}+\sum_{i<j}V_{ij}^{\rm conf}, \label{ham}
\end{equation}
where $H_0$ is a relativistic kinetic energy term
\begin{equation}
H_0 = \sum_{i=1}^{4}(p_i^2+m_i^2)^{1/2}.
\end{equation}
The $V_{ij}^{\rm oge}$ is the one gluon exchange pairwise potential, and $V_{ij}^{\rm conf}$ corresponds to the confining part. The
kinematic energy of the center-of-mass system can be eliminated by the constraint $\sum_{i=1}^4 \boldsymbol p_i = 0$.

In present work, we only concentrate on the $S-$wave ground states and do not include the spin-orbit and tensor interactions.
Then, the potential $V_{ij}^{\rm oge}$ can be expressed as
\begin{equation}
V_{ij}^{\rm{oge}} = \beta_{ij}^{1/2}\tilde G(r_{ij})\beta_{ij}^{1/2} + \delta_{ij}^{1/2+\epsilon_c}\frac{2\boldsymbol{S_i}
\cdot \boldsymbol{S_j}}{3m_im_j} \nabla^2\tilde G(r_{ij})\delta_{ij}^{1/2+\epsilon_c},
\end{equation}
with
\begin{equation}
\beta_{ij} = 1+\frac{p_{ij}^2}{(p_{ij}^2+m_i^2)^{1/2}(p_{ij}^2+m_j^2)^{1/2}},
\end{equation}
and
\begin{equation}
\delta_{ij} = \frac{m_im_j}{(p_{ij}^2+m_i^2)^{1/2}(p_{ij}^2+m_j^2)^{1/2}}.
\end{equation}
The $p_{ij}$ is the magnitude of the momentum of either of the quarks in the center-of-mass frame of $ij$ quark subsystem, and the
$\epsilon_c$ is a free parameter reflecting the momentum dependence. The smeared Coulomb potential $\tilde G(r_{ij})$ is
\begin{equation}
\tilde G(r_{ij}) = \boldsymbol F_i \cdot \boldsymbol F_j \sum_{k=1}^3\frac{\alpha_k}{r_{ij}}{\rm erf}(\tau_{kij}r_{ij}),
\end{equation}
with
\begin{equation}
\frac{1}{\tau_{kij}^2} = \frac{1}{\gamma_k^2}+\frac{1}{\sigma_{ij}^2},
\end{equation}
and
\begin{equation}
\sigma_{ij}^2 = \sigma_0^2
\left[\frac{1}{2}+\frac{1}{2}\left(\frac{4m_im_j}{(m_i+m_j)^2}\right)^4\right]+s^2\left(\frac{2m_im_j}{m_i+m_j}\right)^2.
\end{equation}
The $\boldsymbol F_i \cdot \boldsymbol F_j$ stands for the color matrix and reads
\begin{equation}
\boldsymbol F_i = \left\{
\begin{aligned}
&\frac{\lambda_i}{2}~~~{\rm for~quarks,} \\
&-\frac{\lambda_i^*}{2}~{\rm for~antiquarks.} \\
\end{aligned}
\right.
\end{equation}
Similarly, the confining interaction $V_{ij}^{\rm conf}$ can be expressed as
\begin{equation}
V_{ij}^{\rm{conf}} = -\frac{3}{4}\boldsymbol F_i \cdot \boldsymbol F_j
\left\{ br \left[\frac{e^{-\sigma_{ij}^2r^2}}{\sqrt{\pi}\sigma_{ij}r}+\left(1+\frac{1}{2\sigma_{ij}^2r^2}\right)
{\rm erf}(\sigma_{ij}r)\right]+c\right\}.
\end{equation}
All the parameters used here are taken from the original reference~\cite{Godfrey:1985xj} and collected in
Table~\ref{para} for convenience. The details of the relativized procedure can be found in Refs.~\cite{Godfrey:1985xj,Capstick:1986bm}.

\begin{table}[!htbp]
\begin{center}
\caption{ \label{para} Relevant parameters of the relativized quark model~\cite{Godfrey:1985xj}.}
\begin{tabular*}{8.5cm}{@{\extracolsep{\fill}}*{5}{p{1.25cm}<{\centering}}}
\hline\hline
 $m_u/m_d (\rm MeV)$          &  $m_s (\rm MeV)$            & $m_c (\rm MeV)$       & $m_b (\rm MeV)$   & $\alpha_1$  \\
 220       &  419           & 1628    & 4977       & 0.25  \\\hline
  $\alpha_2$  &  $\alpha_3$   & $\gamma_1 (\rm GeV)$ & $\gamma_2 (\rm GeV)$  & $\gamma_3 (\rm GeV)$ \\
0.15 & 0.20 & 1/2 & $\sqrt{10}/2$ & $\sqrt{1000}/2$ \\\hline
$b (\rm GeV^2)$ & $c (\rm MeV)$     & $\sigma_0 (\rm GeV)$  &  $s$  & $\epsilon_c$ \\
0.18 & -253  & 1.80  & 1.55  &  -0.168  \\
\hline\hline
\end{tabular*}
\end{center}
\end{table}

\subsection{Matrix elements of color, flavor, and spin parts}

The wave function of a $Q_1Q_2^\prime \bar q_3 \bar q_4^\prime$ state can be divided into color, flavor, spin, and spatial parts.
In the color space, one has two kinds of colorless states with well defined permutation properties,
\begin{equation}
|\bar 3 3\rangle = |(Q_1 Q_2^\prime)^{\bar 3} (\bar q_3 \bar q_4^\prime)^3\rangle,
\end{equation}
\begin{equation}
|6 \bar 6\rangle = |(Q_1 Q_2^\prime)^{6} (\bar q_3 \bar q_4^\prime)^{\bar 6}\rangle,
\end{equation}
where the $|\bar 3 3\rangle$ is antisymmetric under the exchange of both quarks and antiquarks, and the $|6 \bar 6 \rangle$ is the
symmetric one. One can evaluate the color matrix elements $\langle \boldsymbol F_i \cdot \boldsymbol F_j \rangle$ with the help of
explicit color wave functions or the SU(3) Casimir operator. The results are collected in Table~\ref{color}.

\begin{table}[!htbp]
\begin{center}
\caption{ \label{color} Color matrix elements.}
\begin{tabular*}{8.5cm}{@{\extracolsep{\fill}}*{7}{p{1.1cm}<{\centering}}}
\hline\hline
$\langle \hat O \rangle$ & $\langle \boldsymbol F_1 \cdot \boldsymbol F_2 \rangle$          &  $\langle \boldsymbol F_3 \cdot
\boldsymbol F_4 \rangle$            & $\langle \boldsymbol F_1 \cdot \boldsymbol F_3 \rangle$   & $\langle \boldsymbol F_2
\cdot \boldsymbol F_4 \rangle$   & $\langle \boldsymbol F_1 \cdot \boldsymbol F_4 \rangle$ & $\langle \boldsymbol F_2
\cdot \boldsymbol F_3 \rangle$ \\\hline
$\langle \bar 3 3 | \hat O |\bar 3 3\rangle$ & -2/3       &  -2/3     &  -1/3   & -1/3   & -1/3 & -1/3  \\
$\langle 6 \bar 6 | \hat O |6 \bar 6 \rangle$ & 1/3       &  1/3     &  -5/6   &  -5/6   &  -5/6 &  -5/6  \\
$\langle \bar 3 3 | \hat O |6 \bar 6\rangle$ & 0       &  0     &  $-1/\sqrt{2}$   & $-1/\sqrt{2}$   & $1/\sqrt{2}$ & $1/\sqrt{2}$  \\
\hline\hline
\end{tabular*}
\end{center}
\end{table}

For the flavor part, the combination between quarks $\bar u$ and $\bar d$ can be symmetric with $I=1$ or antisymmetric with $I=0$, while
the combinations of $\bar s \bar s$, $c c$, and $b b$ are always symmetric. For combinations $\bar u \bar s$ and $\bar d \bar s$, one
can also construct the symmetric and antisymmetric flavor wave functions under the flavor SU(3) symmetry. The $c$ and $b$ are
treated as different particles and no symmetry constraint should be obeyed. For convenience, the notation $\bar u \bar d$ represents
the combinations of  $\bar u \bar u$, $\bar d \bar d$, $(\bar u \bar d+ \bar d \bar u)/\sqrt{2}$, and
$(\bar u \bar d-\bar d\bar u)/\sqrt{2}$, and notation $\bar u \bar s$ stands for the combinations
$(\bar u \bar s+ \bar s \bar u)/\sqrt{2}$, $(\bar u \bar s- \bar s \bar u)/\sqrt{2}$, $(\bar d \bar s+ \bar s \bar d)/\sqrt{2}$,
and $(\bar d \bar s- \bar s \bar d)/\sqrt{2}$ in the present work.

In the spin space, one can construct six spin states,
\begin{equation}
\chi^{00}_0 = |(Q_1 Q_2^\prime)_0 (\bar q_3 \bar q_4^\prime)_0\rangle_0,
\end{equation}
\begin{equation}
\chi^{11}_0 = |(Q_1 Q_2^\prime)_1 (\bar q_3 \bar q_4^\prime)_1\rangle_0,
\end{equation}
\begin{equation}
\chi^{01}_1 = |(Q_1 Q_2^\prime)_0 (\bar q_3 \bar q_4^\prime)_1\rangle_1,
\end{equation}
\begin{equation}
\chi^{10}_1 = |(Q_1 Q_2^\prime)_1 (\bar q_3 \bar q_4^\prime)_0\rangle_1,
\end{equation}
\begin{equation}
\chi^{11}_1 = |(Q_1 Q_2^\prime)_1 (\bar q_3 \bar q_4^\prime)_1\rangle_1,
\end{equation}
\begin{equation}
\chi^{11}_2 = |(Q_1 Q_2^\prime)_1 (\bar q_3 \bar q_4^\prime)_1\rangle_2,
\end{equation}
where $(Q_1 Q_2^\prime)_0$ and $(\bar q_3 \bar q_4^\prime)_0$ are antisymmetric for the two fermions under permutations, and the
$(Q_1 Q_2^\prime)_1$ and $(\bar q_3 \bar q_4^\prime)_1$ are symmetric. For the notation $\chi^{S_{12} S_{34}}_S$, the $S_{12}$,
$S_{34}$, and $S$ are the spin of two heavy quarks, spin of two light antiquarks, and total spin, respectively. The relevant spin
matrix elements can be evaluated with the standard angular momentum algebra, and the results are listed in Table~\ref{spin}.

\begin{table}[!htbp]
\begin{center}
\caption{ \label{spin} Spin matrix elements.}
\begin{tabular*}{8.5cm}{@{\extracolsep{\fill}}*{7}{p{1.1cm}<{\centering}}}
\hline\hline
$\langle \hat O \rangle$ & $\langle \boldsymbol S_1 \cdot \boldsymbol S_2 \rangle$          &  $\langle \boldsymbol S_3 \cdot
\boldsymbol S_4 \rangle$            & $\langle \boldsymbol S_1 \cdot \boldsymbol S_3 \rangle$   & $\langle \boldsymbol S_2
\cdot \boldsymbol S_4 \rangle$   & $\langle \boldsymbol S_1 \cdot \boldsymbol S_4 \rangle$ & $\langle \boldsymbol S_2 \cdot
\boldsymbol S_3 \rangle$ \\\hline
$\langle \chi^{00}_0 | \hat O |\chi^{00}_0 \rangle$ & -3/4       &  -3/4     &  0   & 0   & 0 & 0  \\
$\langle \chi^{11}_0 | \hat O |\chi^{11}_0 \rangle$ & 1/4       &  1/4     &  -1/2   & -1/2   & -1/2 & -1/2  \\
$\langle \chi^{00}_0 | \hat O |\chi^{11}_0 \rangle$ & 0       &  0     &  $-\sqrt{3}/4$   & $-\sqrt{3}/4$   & $\sqrt{3}/4$
& $\sqrt{3}/4$  \\
$\langle \chi^{01}_1 | \hat O |\chi^{01}_1 \rangle$ & -3/4       &  1/4     &  0   & 0   & 0 & 0  \\
$\langle \chi^{10}_1 | \hat O |\chi^{10}_1 \rangle$ & 1/4       &  -3/4     &  0   & 0   & 0 & 0  \\
$\langle \chi^{11}_1 | \hat O |\chi^{11}_1 \rangle$ & 1/4       &  1/4     &  -1/4   & -1/4   & -1/4 & -1/4  \\
$\langle \chi^{01}_1 | \hat O |\chi^{10}_1 \rangle$ & 0       & 0     &  1/4   & 1/4   & -1/4 & -1/4  \\
$\langle \chi^{01}_1 | \hat O |\chi^{11}_1 \rangle$ & 0       & 0     &  $-\sqrt{2}/4$   & $\sqrt{2}/4$   & $-\sqrt{2}/4$
& $\sqrt{2}/4$  \\
$\langle \chi^{10}_1 | \hat O |\chi^{11}_1 \rangle$ & 0       & 0     &  $\sqrt{2}/4$   & $-\sqrt{2}/4$   & $-\sqrt{2}/4$
& $\sqrt{2}/4$  \\
$\langle \chi^{11}_2 | \hat O |\chi^{11}_2 \rangle$ & 1/4       & 1/4     &  1/4   & 1/4   & 1/4 & 1/4  \\
\hline\hline
\end{tabular*}
\end{center}
\end{table}

For a $S-$wave $T_{QQ^\prime}$ state, the spatial part is always symmetric, and then the color-spin-flavor wave function should be
antisymmetric for the identical quarks and antiquarks according to the Pauli exclusion principle. From the above discussions, we
perform all possible configurations for the $QQ^\prime \bar q \bar q^\prime$ systems in Table~\ref{configuration}. It should be noted
that for a given system different configurations with same isospin-spin can mix with each other.

\begin{table*}[!htbp]
\begin{center}
\caption{ \label{configuration} All possible configurations for the $QQ^\prime \bar q \bar q^\prime$ systems.
The subscripts and superscripts are the
spin quantum numbers and color types, respectively. The braces $\{ ~\}$, brackets $[~]$ strand for the symmetric, antisymmetric
flavor wave functions, respectively. The parentheses $(~)$ are used for the subsystems without permutation symmetries.}
\begin{tabular*}{18cm}{@{\extracolsep{\fill}}*{5}{p{3cm}<{\centering}}}
\hline\hline
System & $IJ^P$         &  \multicolumn{3}{c}{Configuration} \\\hline
$\{ c c\} [\bar u \bar d] $ & $01^+$       &  $|\{cc\}^{\bar 3}_1 [\bar u \bar d]^3_0\rangle_1$
&  $|\{cc\}^6_0 [\bar u \bar d]^{\bar 6}_1\rangle_1$   & $\cdots$     \\
$\{ c c\} \{\bar u \bar d\}$ & $10^+$       &  $|\{cc\}^{\bar 3}_1 \{\bar u \bar d\}^3_1\rangle_0$
&  $|\{cc\}^6_0 \{\bar u \bar d\}^{\bar 6}_0\rangle_0$     & $\cdots$  \\
& $11^+$       &  $|\{cc\}^{\bar 3}_1 \{\bar u \bar d\}^3_1\rangle_1$     &  $\cdots$   & $\cdots$   \\
& $12^+$       &  $|\{cc\}^{\bar 3}_1 \{\bar u \bar d\}^3_1\rangle_2$     &  $\cdots$   & $\cdots$    \\
$\{bb\}[\bar u \bar d] $ & $01^+$       &  $|\{bb\}^{\bar 3}_1 [\bar u \bar d]^3_0\rangle_1$
&  $|\{bb\}^6_0 [\bar u \bar d]^{\bar 6}_1\rangle_1$   & $\cdots$   \\
$\{bb\}\{\bar u \bar d\} $& $10^+$       &  $|\{bb\}^{\bar 3}_1 \{\bar u \bar d\}^3_1\rangle_0$
&  $|\{bb\}^6_0 \{\bar u \bar d\}^{\bar 6}_0\rangle_0$      & $\cdots$  \\
& $11^+$       &  $|\{bb\}^{\bar 3}_1 \{\bar u \bar d\}^3_1\rangle_1$      &  $\cdots$   & $\cdots$    \\
& $12^+$       &  $|\{bb\}^{\bar 3}_1 \{\bar u \bar d\}^3_1\rangle_2$     &  $\cdots$   & $\cdots$   \\
$(cb)[\bar u \bar d] $ & $00^+$ &  $|(cb)^{\bar 3}_0 [\bar u \bar d]^3_0 \rangle_0$
&  $|(cb)^6_1 [\bar u \bar d]^{\bar 6}_1\rangle_0$     & $\cdots$\\
& $01^+$       &  $|(cb)^{\bar 3}_1 [\bar u \bar d]^3_0\rangle_1$     &  $|(cb)^6_0 [\bar u \bar d]^{\bar 6}_1\rangle_1$
&  $|(cb)^6_1 [\bar u \bar d]^{\bar 6}_1\rangle_1$    \\
& $02^+$       &   $|(cb)^6_1 [\bar u \bar d]^{\bar 6}_1\rangle_2$   & $\cdots$  & $\cdots$    \\
$(cb)\{\bar u \bar d\} $& $10^+$       & $|(cb)^{\bar 3}_1 \{\bar u \bar d\}^3_1\rangle_0$
&  $|(cb)^6_0 \{\bar u \bar d\}^{\bar 6}_0\rangle_0$   & $\cdots$    \\
& $11^+$      &  $|(cb)^{\bar 3}_0 \{\bar u \bar d\}^3_1\rangle_1$  &  $|(cb)^{\bar 3}_1 \{\bar u \bar d\}^3_1\rangle_1$
&   $|(cb)^6_1 \{\bar u \bar d\}^{\bar 6}_0\rangle_1$     \\
& $12^+$       &  $|(cb)^{\bar 3}_1 \{\bar u \bar d\}^3_1\rangle_2$     &  $\cdots$   & $\cdots$     \\\hline

$\{cc\}[\bar u \bar s] $ & $\frac{1}{2}1^+$       &  $|\{cc\}^{\bar 3}_1 [\bar u \bar s]^3_0\rangle_1$
&  $|\{cc\}^6_0 [\bar u \bar s]^{\bar 6}_1\rangle_1$   & $\cdots$     \\
$\{cc\}\{\bar u \bar s\} $& $\frac{1}{2}0^+$       &  $|\{cc\}^{\bar 3}_1 \{\bar u \bar s\}^3_1\rangle_0$
 &  $|\{cc\}^6_0 \{\bar u \bar s\}^{\bar 6}_0\rangle_0$     & $\cdots$  \\
& $\frac{1}{2}1^+$       &  $|\{cc\}^{\bar 3}_1 \{\bar u \bar s\}^3_1\rangle_1$     &  $\cdots$   & $\cdots$   \\
& $\frac{1}{2}2^+$        &  $|\{cc\}^{\bar 3}_1 \{\bar u \bar s\}^3_1\rangle_2$     &  $\cdots$   & $\cdots$    \\
$\{bb\}[\bar u \bar s] $ & $\frac{1}{2}1^+$       &  $|\{bb\}^{\bar 3}_1 [\bar u \bar s]^3_0\rangle_1$
 &  $|\{bb\}^6_0 [\bar u \bar s]^{\bar 6}_1\rangle_1$   & $\cdots$   \\
$\{bb\}\{\bar u \bar s\} $& $\frac{1}{2}0^+$      &  $|\{bb\}^{\bar 3}_1 \{\bar u \bar s\}^3_1\rangle_0$
 &  $|\{bb\}^6_0 \{\bar u \bar s\}^{\bar 6}_0\rangle_0$      & $\cdots$  \\
& $\frac{1}{2}1^+$        &  $|\{bb\}^{\bar 3}_1 \{\bar u \bar s\}^3_1\rangle_1$      &  $\cdots$   & $\cdots$    \\
& $\frac{1}{2}2^+$       &  $|\{bb\}^{\bar 3}_1 \{\bar u \bar s\}^3_1\rangle_2$     &  $\cdots$   & $\cdots$   \\
$(cb)[\bar u \bar s] $ & $\frac{1}{2}0^+$ &  $|(cb)^{\bar 3}_0 [\bar u \bar s]^3_0 \rangle_0$
&  $|(cb)^6_1 [\bar u \bar s]^{\bar 6}_1\rangle_0$     & $\cdots$\\
& $\frac{1}{2}1^+$       &  $|(cb)^{\bar 3}_1 [\bar u \bar s]^3_0\rangle_1$
 &  $|(cb)^6_0 [\bar u \bar s]^{\bar 6}_1\rangle_1$   &  $|(cb)^6_1 [\bar u \bar s]^{\bar 6}_1\rangle_1$    \\
& $\frac{1}{2}2^+$       &   $|(cb)^6_1 [\bar u \bar s]^{\bar 6}_1\rangle_2$   & $\cdots$  & $\cdots$    \\
$(cb)\{\bar u \bar s\} $  & $\frac{1}{2}0^+$ & $|(cb)^{\bar 3}_1 \{\bar u \bar s\}^3_1\rangle_0$
 &  $|(cb)^6_0 \{\bar u \bar s\}^{\bar 6}_0\rangle_0$   & $\cdots$    \\
& $\frac{1}{2}1^+$      &  $|(cb)^{\bar 3}_0 \{\bar u \bar s\}^3_1\rangle_1$  &  $|(cb)^{\bar 3}_1 \{\bar u \bar s\}^3_1\rangle_1$
&   $|(cb)^6_1 \{\bar u \bar s\}^{\bar 6}_0\rangle_1$     \\
& $\frac{1}{2}2^+$       & $|(cb)^{\bar 3}_1 \{\bar u \bar s\}^3_1\rangle_2$     &  $\cdots$
& $\cdots$     \\\hline

$ \{cc\} \{\bar s \bar s\}$ & $00^+$       &  $|\{cc\}^{\bar 3}_1 \{\bar s \bar s\}^3_1\rangle_0$
&  $|\{cc\}^6_0 \{\bar s \bar s\}^{\bar 6}_0\rangle_0$     & $\cdots$  \\
& $01^+$       &  $|\{cc\}^{\bar 3}_1 \{\bar s \bar s\}^3_1\rangle_1$     &  $\cdots$   & $\cdots$   \\
& $02^+$       &  $|\{cc\}^{\bar 3}_1 \{\bar s \bar s\}^3_1\rangle_2$     &  $\cdots$   & $\cdots$    \\
$\{bb\}\{\bar s \bar s\} $ & $00^+$       &  $|\{bb\}^{\bar 3}_1 \{\bar s \bar s\}^3_1\rangle_0$
 &  $|\{bb\}^6_0 \{\bar s \bar s\}^{\bar 6}_0\rangle_0$      & $\cdots$  \\
& $01^+$       &  $|\{bb\}^{\bar 3}_1 \{\bar s \bar s\}^3_1\rangle_1$      &  $\cdots$   & $\cdots$    \\
& $02^+$       &  $|\{bb\}^{\bar 3}_1 \{\bar s \bar s\}^3_1\rangle_2$     &  $\cdots$   & $\cdots$   \\
$(cb)\{\bar s \bar s\}$ & $00^+$       & $|(cb)^{\bar 3}_1 \{\bar s \bar s\}^3_1\rangle_0$
&  $|(cb)^6_0 \{\bar s \bar s\}^{\bar 6}_0\rangle_0$   & $\cdots$    \\
& $01^+$      &  $|(cb)^{\bar 3}_0 \{\bar s \bar s\}^3_1\rangle_1$  &  $|(cb)^{\bar 3}_1 \{\bar s \bar s\}^3_1\rangle_1$
&   $|(cb)^6_1 \{\bar s \bar s\}^{\bar 6}_0\rangle_1$     \\
& $02^+$       & $|(cb)^{\bar 3}_1 \{\bar s \bar s\}^3_1\rangle_2$     &  $\cdots$   & $\cdots$     \\\hline
\end{tabular*}
\end{center}
\end{table*}

\subsection{Matrix elements of spatial part}

\begin{figure*}[!htbp]
\includegraphics[scale=0.7]{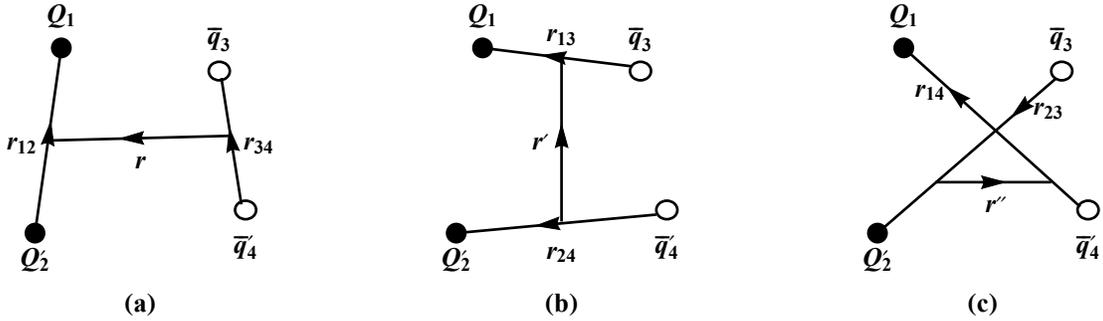}
\vspace{0.0cm} \caption{The $Q_1Q_2^\prime \bar q_3 \bar q_4^\prime$ tetraquark state in Jacobi coordinates.}
\label{jacobi}
\end{figure*}

For a $Q_1Q_2^\prime \bar q_3 \bar q_4^\prime$ state, the Jacobi coordinates are shown in Figure~\ref{jacobi}. In these coordinates,
one can define
\begin{equation}
\boldsymbol r_{12}=  \boldsymbol r_1 - \boldsymbol r_2,
\end{equation}
\begin{equation}
\boldsymbol r_{34}=  \boldsymbol r_3 - \boldsymbol r_4,
\end{equation}
\begin{equation}
\boldsymbol r = \frac{m_1 \boldsymbol r_1 + m_2 \boldsymbol r_2}{m_1+m_2} - \frac{m_3 \boldsymbol r_3 + m_4 \boldsymbol r_4}{m_3+m_4},
\end{equation}
and
\begin{equation}
\boldsymbol R = \frac{m_1 \boldsymbol r_1 + m_2 \boldsymbol r_2
+ m_3 \boldsymbol r_3 + m_4 \boldsymbol r_4}{m_1+m_2+m_3+m_4}.
\end{equation}
Then, other relevant coordinates of this system can be expressed in terms of $\boldsymbol r_{12}$, $\boldsymbol r_{34}$,
and $\boldsymbol r$ as follows
\begin{equation}
\boldsymbol r_{13} = \boldsymbol r_1 - \boldsymbol r_3 = \frac{m_2 }{m_1+m_2}\boldsymbol r_{12}
- \frac{m_4}{m_3+m_4} \boldsymbol r_{34} + \boldsymbol r,
\end{equation}
\begin{equation}
\boldsymbol r_{24} = \boldsymbol r_2 - \boldsymbol r_4 = -\frac{m_1 }{m_1+m_2} \boldsymbol r_{12}
+ \frac{m_3}{m_3+m_4} \boldsymbol r_{34} + \boldsymbol r,
\end{equation}
\begin{equation}
\boldsymbol r_{14} = \boldsymbol r_1 - \boldsymbol r_4 = \frac{m_2}{m_1+m_2} \boldsymbol r_{12}
+ \frac{m_3}{m_3+m_4} \boldsymbol r_{34} + \boldsymbol r,
\end{equation}
\begin{equation}
\boldsymbol r_{23} = \boldsymbol r_2 - \boldsymbol r_3 = -\frac{m_1}{m_1+m_2} \boldsymbol r_{12}
- \frac{m_4}{m_3+m_4} \boldsymbol r_{34} + \boldsymbol r,
\end{equation}
\begin{eqnarray}
\boldsymbol r^\prime &=& \frac{m_1 \boldsymbol r_1 + m_3 \boldsymbol r_3}{m_1+m_3}
- \frac{m_2 \boldsymbol r_2 + m_4 \boldsymbol r_4}{m_2+m_4}\nonumber\\
 &=& \frac{m_1m_2(m_1+m_2+m_3+m_4)}{(m_1+m_2)(m_1+m_3)(m_2+m_4)} \boldsymbol r_{12}
 + \nonumber\\ && \frac{m_3m_4(m_1+m_2+m_3+m_4)}{(m_3+m_4)(m_1+m_3)(m_2+m_4)}
 \boldsymbol r_{34} + \nonumber\\ &&  \frac{m_1m_4-m_2m_3}{(m_1+m_3)(m_2+m_4)} \boldsymbol r,
\end{eqnarray}
\begin{eqnarray}
\boldsymbol r^{\prime\prime} &=& \frac{m_1 \boldsymbol r_1 + m_4 \boldsymbol r_4}{m_1+m_4}
- \frac{m_2 \boldsymbol r_2 + m_3 \boldsymbol r_3}{m_2+m_3}\nonumber\\
 &=& \frac{m_1m_2(m_1+m_2+m_3+m_4)}{(m_1+m_2)(m_1+m_4)(m_2+m_3)}
 \boldsymbol r_{12}
 - \nonumber\\ && \frac{m_3m_4(m_1+m_2+m_3+m_4)}{(m_3+m_4)(m_1+m_4)(m_2+m_3)}
 \boldsymbol r_{34} + \nonumber\\ && \frac{m_1m_3-m_2m_4}{(m_1+m_4)(m_2+m_3)} \boldsymbol r.
\end{eqnarray}

In our numerical calculation, the spatial wave function of a few-body system can be expanded in terms of a set of Gaussian basis
functions, which forms an approximate complete set in a finite coordinate space~\cite{Hiyama:2003cu}. For a $S-$wave
$Q_1Q_2^\prime \bar q_3 \bar q_4^\prime$ tetraquark, the expanded basis should satisfy the relation
$\boldsymbol l_{12} + \boldsymbol l_{34} +\boldsymbol l=0$, where the $\boldsymbol l_{12}$, $\boldsymbol l_{34}$, and $\boldsymbol l$
are the relative angular momenta of the  $Q_1 Q_2^\prime$, $\bar q_3 \bar q_4^\prime$, and $(Q_1 Q_2^\prime)(\bar q_3 \bar q_4^\prime)$,
respectively. The contributions of higher orbital excitations to the ground states arise from the slight mixing via the spin-orbit
or tensor interactions, which have been neglected in present calculations. Then, only the $l_{12}= l_{34}= l=0$ case should be
considered, and the spatial wave function for a certain tetraquark configuration can be expressed as
\begin{equation}
\Psi(\boldsymbol r_{12},\boldsymbol r_{34},\boldsymbol r) = \sum_{n_Q,n_q,n} C_{n_Qn_qn}
\psi_{n_Q}(\boldsymbol r_{12}) \psi_{n_q}(\boldsymbol r_{34}) \psi_n(\boldsymbol r),
\end{equation}
where $C_{n_Qn_qn}$ are the expansion coefficients. The  $\psi_{n_Q}(\boldsymbol r_{12}) \psi_{n_q}(\boldsymbol r_{34})
\psi_n(\boldsymbol r)$ stands for the position representation of the basis $|\alpha \rangle \equiv |n_Qn_qn\rangle$, where
\begin{equation}
\psi_n(\boldsymbol r) = \frac{2^{7/4}\nu_n^{3/4}}{\pi^{1/4}} e^{-\nu_n r^2} Y_{00}(\hat{\boldsymbol r}) =
\Bigg(\frac{2 \nu_n}{\pi} \Bigg )^{3/4} e^{-\nu_n r^2},
\end{equation}
\begin{equation}
\nu_n = \frac{1}{r_1^2a^{2(n-1)}},~~~~ (n=1-N_{max}).
\end{equation}
The three parameters $r_1$, $a$, and $N_{max}$ are the Gaussian size parameters in geometric progression for numerical calculations,
and the final results are stable and independent with these parameters within an approximate complete set in a sufficiently large
space~\cite{Hiyama:2003cu}. Besides the position representation $\psi_n(\boldsymbol r)$, it is also convenient for the numerical
calculations to present the momentum representation $\phi_n(\boldsymbol p)$,
\begin{equation}
\phi_n(\boldsymbol p) = \frac{2^{1/4}}{\pi^{1/4}\nu_n^{3/4}} e^{- p^2/(4\nu_n)} Y_{00}(\hat{\boldsymbol p}) = \Bigg(\frac{1}{2
\pi \nu_n } \Bigg )^{3/4} e^{- p^2/(4\nu_n)}.
\end{equation}
Similarly, the formulas of $\psi_{n_Q}(\boldsymbol r_{12})$, $\phi_{n_Q}(\boldsymbol p_{12})$, $\psi_{n_q}(\boldsymbol r_{34})$,
and $\phi_{n_q}(\boldsymbol p_{34})$ can be obtained by replacing the $n$, $\boldsymbol r$, and $\boldsymbol p$ of the
$\psi_n(\boldsymbol r)$ and $\phi_n(\boldsymbol p)$.

To calculate the spatial matrix elements, we encounter the momentum-dependent factors combined with the position-dependent potentials
in the relativized Hamiltonian. This difficulty can be overcomed by inserting complete sets of Guassian functions between the two
types of operators. Take the first term of $V_{ij}^{oge}$ for example, the matrix elements between two bases $|\alpha \rangle$ and
$|\beta \rangle$ can be written as
\begin{eqnarray}
\langle \alpha |\beta_{ij}^{1/2}\tilde G(r_{ij})\beta_{ij}^{1/2} | \beta \rangle &=& \sum_{\gamma,\delta,\rho,\lambda} \langle
\alpha |\beta_{ij}^{1/2}  | \gamma \rangle (N^{-1})_{\gamma\delta} \langle \delta | \tilde G(r_{ij}) | \rho \rangle \nonumber \\
&& \times (N^{-1})_{\rho\lambda} \langle \lambda |\beta_{ij}^{1/2} | \beta \rangle.
\end{eqnarray}
The $N$ is the overlap matrix of the Guassian functions with matrix elements $N_{ij} = \langle i | j \rangle$, which arises form
the nonorthogonality of the bases. Together with the explicit forms of the basis in two representations, one can evaluate the
expectations of momentum-dependent parts and position-dependent parts in the momentum representation and position representation,
respectively.

\subsection{Generalized eigenvalue problem}

When all the matrix elements have been worked out, the mass spectra can be obtained by solving the generalized eigenvalue problem.
For a given configuration without mixing, the homogeneous equation set can be expressed as
\begin{equation}
\sum_{j=1}^{N_{max}^3}(H_{ij}-EN_{ij})C_j=0,~~~~ (i=1-N_{max}^3).
\end{equation}
Where, the $H_{ij}$ are the matrix elements in the total color-flavor-spin-spatial bases, $E$ stands for the eigenvalue, and $C_j$
are the relevant eigenvector. The lowest eigenvalue represents for the mass of this configuration, and the eigenvector corresponds
to the  expansion coefficients $C_{n_Qn_qn}$ in the spatial wave function.

From Table~\ref{configuration}, a given system may include several different configurations with same $IJ^P$, which can mix with
each other. In present calculation, we first solve the generalized eigenvalue problem to get the masses of pure configurations,
and then calculate the off-diagonal effects between different configurations. The final mass spectra can be obtained by diagonalizing
the mass matrix of these configurations.

\section{RESULTS AND DISCUSSIONS}{\label{results}}

\subsection{Numerical stability}

Before discussing the properties of predicted tetraquarks, It is important to concentrate on the stabilities of the numerical
procedures. In the nonrelativistic quark model, one can calculate the expectations of Hamiltonian in the trial wave functions,
and always obtain the upper limit of the masses. When the number of bases increases, the numerical results decrease and approximate
closely to the actual values. Empirically, stable results for $S-$ wave states can be achieved within small numbers of bases.

In the relativized quark model, to calculate the matrix elements of Hamiltonian, complete sets of Guassian functions should be
inserted twice for the $V_{ij}^{oge}$, while the $V_{ij}^{conf}$ and relativistic kinetic energy term can be evaluated straightforward.
The number of basis should be large enough to guarantee approximate completeness, otherwise the matrix elements of $V_{ij}^{oge}$
terms will be meaningless. For the meson spectra, a dozen bases are adequate, while about one hundred bases are needed for the
baryon spectra~\cite{Godfrey:1985xj,Capstick:1986bm}. One can expect that several hundred or one thousand Guassian functions are
proper for calculating the tetraquark spectra.

\begin{figure}[!htbp]
\includegraphics[scale=0.7]{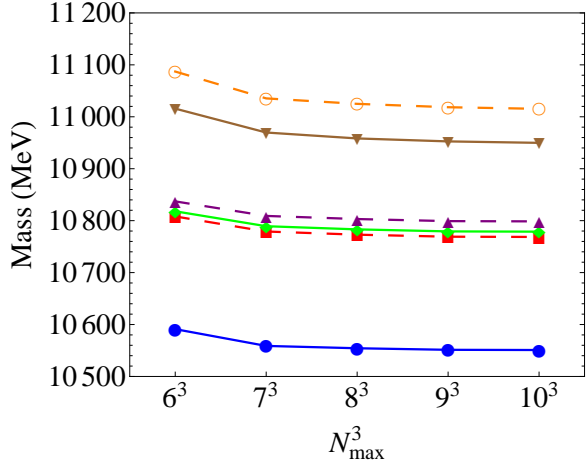}
\vspace{0.0cm} \caption{Numerical stabilities for six pure configurations of $bb \bar u \bar d$ system. The blue points, red squares,
green diamonds, purple triangles, brown inverted triangles, and orange circles stand for the $|\{bb\}^{\bar 3}_1 [\bar u
\bar d]^3_0\rangle_1$, $|\{bb\}^{\bar 3}_1 \{\bar u \bar d\}^3_1\rangle_0$, $|\{bb\}^{\bar 3}_1 \{\bar u \bar d\}^3_1\rangle_1$,
$|\{bb\}^{\bar 3}_1 \{\bar u \bar d\}^3_1\rangle_2$, $|\{bb\}^6_0 [\bar u \bar d]^{\bar 6}_1\rangle_1$, and $|\{bb\}^6_0 \{\bar u
\bar d\}^{\bar 6}_0\rangle_0$ configurations.}
\label{stability}
\end{figure}

Take the six pure configurations of $bb \bar u \bar d$ system for example, we investigate the dependence of results on the number
of bases. The basis number varies from $N^3_{max} = 6^3$ to $10^3$, and the dependence is presented in Figure~\ref{stability}. It is
found that the eigenvalues are stable when the $N^3_{max}$ becomes larger. With $N^3_{max} = 10^3$ bases, the numerical uncertainties
are rather small, which are enough for the quark model calculations. Thus, we adopt $10^3$ Gaussian bases to study the
$S-$wave $T_{QQ^\prime}$ spectra in present work.

\subsection{Non-strange systems}

\begin{figure*}[!htbp]
\includegraphics[scale=0.5]{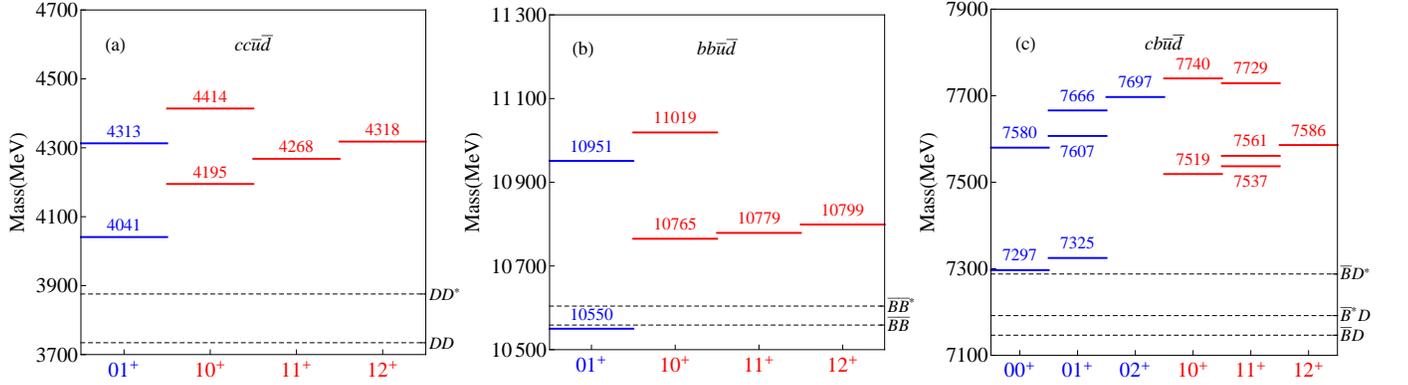}
\vspace{-0.5cm} \caption{The predicted masses of $cc\bar u \bar d$, $bb\bar u \bar d$, and $cb\bar u \bar d$ systems together with
relevant thresholds. The blue lines stand for the tetraquarks including antisymmetric light subsystem $[\bar u \bar d]$, and the red
lines correspond to the ones with symmetric light subsystem $\{\bar u \bar d\}$.}
\label{ud}
\end{figure*}

The predicted masses of $cc\bar u \bar d$, $bb\bar u \bar d$, and $cb\bar u \bar d$ systems are presented in Table~\ref{mass1}
and Figure~\ref{ud}. For the $cc\bar u \bar d$ system, the lowest state is the $IJ^P=01^+$ one with 4041 MeV, which is a mixing
state of the $|\{cc\}^{\bar 3}_1 [\bar u \bar d]^3_0\rangle_1$ and $|\{ cc \}^6_0 [\bar u \bar d ]^{\bar 6}_1\rangle_1$ configurations.
This mixing is relatively small, and the $|\{cc\}^{\bar 3}_1 [\bar u \bar d]^3_0\rangle_1$ component is predominant. Due to the quantum
conservation, the $0^+$ and $2^+$ states might decay into a pair of pseudoscalar mesons, while the allowed decay mode of a $1^+$ state
should be a vector meson plus a pseudoscalar one. From Figure~\ref{ud}, it can be seen that the lowest $cc\bar u \bar d$ state is 165 MeV
higher than the $DD^*$ threshold, which can easily decay via falling apart mechanism.

\begin{table*}[htp]
\begin{center}
\caption{\label{mass1} Predicted mass spectra for the $cc\bar u \bar d$, $bb\bar u \bar d$, and $cb\bar u \bar d$ systems.}
\begin{tabular*}{18cm}{@{\extracolsep{\fill}}*{5}{p{3.3cm}<{\centering}}}
\hline\hline
 $IJ^P$  & Configuration                                             & $\langle H\rangle$ (MeV) & Mass (MeV)  & Eigenvector\\\hline
 $01^+$   & $|\{cc\}^{\bar 3}_1 [\bar u \bar d]^3_0\rangle_1$       & \multirow{2}{*}{$\begin{pmatrix}4053&-55 \\-55&4302\end{pmatrix}$}
               & \multirow{2}{*}{$\begin{bmatrix}4041 \\4313 \end{bmatrix}$}  & \multirow{2}{*}{$\begin{bmatrix}(-0.979, -0.205)\\
               (0.205, -0.979)\end{bmatrix}$}\\
                & $|\{cc\}^6_0 [\bar u \bar d]^{\bar 6}_1\rangle_1$ \\
 $10^+$  &  $|\{cc\}^{\bar 3}_1 \{\bar u \bar d\}^3_1\rangle_0$     & \multirow{2}{*}{$\begin{pmatrix}4241&-89 \\-89&4369\end{pmatrix}$}
               & \multirow{2}{*}{$\begin{bmatrix}4195 \\4414 \end{bmatrix}$}  & \multirow{2}{*}{$\begin{bmatrix}(-0.890, -0.455)\\
               (0.455, -0.890)\end{bmatrix}$}\\
                 &  $|\{cc\}^6_0 \{\bar u \bar d\}^{\bar 6}_0\rangle_0$    \\
 $11^+$  &  $|\{cc\}^{\bar 3}_1 \{\bar u \bar d\}^3_1\rangle_1$    & 4268  &  4268  &  1\\
 $12^+$  &  $|\{cc\}^{\bar 3}_1 \{\bar u \bar d\}^3_1\rangle_2$    & 4318  &  4318  &  1\\\hline

 $01^+$   & $|\{bb\}^{\bar 3}_1 [\bar u \bar d]^3_0\rangle_1$       & \multirow{2}{*}{$\begin{pmatrix}10551&20 \\20&10950\end{pmatrix}$}
               & \multirow{2}{*}{$\begin{bmatrix}10550 \\10951 \end{bmatrix}$}  & \multirow{2}{*}{$\begin{bmatrix}(-0.999, 0.050)\\
               (-0.050, -0.999)\end{bmatrix}$}\\
                & $|\{bb\}^6_0 [\bar u \bar d]^{\bar 6}_1\rangle_1$ \\
 $10^+$  &  $|\{bb\}^{\bar 3}_1 \{\bar u \bar d\}^3_1\rangle_0$     & \multirow{2}{*}{$\begin{pmatrix}10769&31 \\31&11015\end{pmatrix}$}
               & \multirow{2}{*}{$\begin{bmatrix}10765 \\11019 \end{bmatrix}$}  & \multirow{2}{*}{$\begin{bmatrix}(-0.993, 0.122)\\
               (-0.122, -0.993)\end{bmatrix}$}\\
                 &  $|\{bb\}^6_0 \{\bar u \bar d\}^{\bar 6}_0\rangle_0$    \\
 $11^+$  &  $|\{bb\}^{\bar 3}_1 \{\bar u \bar d\}^3_1\rangle_1$    & 10779  &  10779  &  1\\
 $12^+$  &  $|\{bb\}^{\bar 3}_1 \{\bar u \bar d\}^3_1\rangle_2$    & 10799  &  10799  &  1\\\hline

 $00^+$  &  $|(cb)^{\bar 3}_0 [\bar u \bar d]^3_0 \rangle_0$      & \multirow{2}{*}{$\begin{pmatrix}7314&-67 \\-67&7563\end{pmatrix}$}
               & \multirow{2}{*}{$\begin{bmatrix}7297 \\7580 \end{bmatrix}$}  & \multirow{2}{*}{$\begin{bmatrix}(-0.970, -0.245)\\
               (0.245, -0.970)\end{bmatrix}$}\\
               &  $|(cb)^6_1 [\bar u \bar d]^{\bar 6}_1\rangle_0$ \\
 $01^+$ &  $|(cb)^{\bar 3}_1 [\bar u \bar d]^3_0\rangle_1$      & \multirow{3}{*}{$\begin{pmatrix}7330&-35&17 \\-35&7658&18 \\
 17& 18 & 7611\end{pmatrix}$}
               & \multirow{3}{*}{$\begin{bmatrix}7325 \\7607 \\ 7666 \end{bmatrix}$}  & \multirow{3}{*}{$\begin{bmatrix}
               (-0.992, -0.109, 0.067)\\(0.095, -0.274, 0.957) \\ (-0.086, 0.956, 0.282)\end{bmatrix}$}\\
                &  $|(cb)^6_0 [\bar u \bar d]^{\bar 6}_1\rangle_1$     \\
              &  $|(cb)^6_1 [\bar u \bar d]^{\bar 6}_1\rangle_1$ \\
$02^+$ &   $|(cb)^6_1 [\bar u \bar d]^{\bar 6}_1\rangle_2$    & 7697  &  7697  &  1\\

 $10^+$  & $|(cb)^{\bar 3}_1 \{\bar u \bar d\}^3_1\rangle_0$    & \multirow{2}{*}{$\begin{pmatrix}7535&-56 \\-56&7724\end{pmatrix}$}
               & \multirow{2}{*}{$\begin{bmatrix}7519 \\7740 \end{bmatrix}$}  & \multirow{2}{*}{$\begin{bmatrix}(-0.964, -0.265)\\
               (0.265, -0.964)\end{bmatrix}$}\\
                 &   $|(cb)^6_0 \{\bar u \bar d\}^{\bar 6}_0\rangle_0$    \\
 $11^+$  &  $|(cb)^{\bar 3}_0 \{\bar u \bar d\}^3_1\rangle_1$     & \multirow{3}{*}{$\begin{pmatrix}7553&10&32 \\10&7552&-16 \\
 32& -16 & 7722\end{pmatrix}$}
               & \multirow{3}{*}{$\begin{bmatrix}7537 \\7561 \\ 7729 \end{bmatrix}$}  & \multirow{3}{*}{$\begin{bmatrix}(-0.740,
               0.648, 0.183)\\(-0.650, -0.758, 0.054) \\ (-0.174, 0.079, -0.982)\end{bmatrix}$}\\
                &  $|(cb)^{\bar 3}_1 \{\bar u \bar d\}^3_1\rangle_1$     \\
              &    $|(cb)^6_1 \{\bar u \bar d\}^{\bar 6}_0\rangle_1$    \\
$12^+$    &  $|(cb)^{\bar 3}_1 \{\bar u \bar d\}^3_1\rangle_2$    & 7586  &  7586  &  1\\

\hline\hline
\end{tabular*}
\end{center}
\end{table*}

For the $bb\bar u \bar d$ system, the mixing between different configurations are rather small and can be neglected. The predicted
mass of the lowest state is 10550 MeV, which is almost a pure $|\{bb\}^{\bar 3}_1 [\bar u \bar d]^3_0\rangle_1$ state. From our
calculation, its mass is lower than the $\bar B \bar B$ and $\bar B \bar B^*$ thresholds, which indicates that both strong
and electromagnetic decays are forbidden. Compared with $\bar B \bar B^*$ channel, the binding energy is 54 MeV and the decay width
should be tiny enough. Although the binding energy is smaller than that of the nonrelativistic quark
models~\cite{Semay:1994ht,Pepin:1996id,Brink:1998as,Vijande:2003ki,Zouzou:1986qh,Park:2018wjk,Deng:2018kly,Hernandez:2019eox,
Yang:2019itm,Maiani:2019cwl,Maiani:2019lpu}, we obtain the same conclusion about the stability of this state. The differences may arise
from the relativized Hamiltonian, where the smearing potentials and relativistic corrections are included. This narrow structure
can be searched via final states of weak decays, such as $\bar B D \pi^-$ and $\bar B D l^- \nu_l$, in future LHC experiments~\cite{Bediaga:2018lhg,Cerri:2018ypt}.

For the  $cb\bar u \bar d$ system, there are two lower states around 7.3 GeV. With small mixing, these two states mainly consist of
$|(cb)^{\bar 3}_0 [\bar u \bar d]^3_0 \rangle_0$ and $|(cb)^{\bar 3}_1 [\bar u \bar d]^3_0 \rangle_1$ configurations, respectively.
The predicted masses of $cb \bar u \bar d$ tetraquarks are much higher than the $D \bar B$ and $D \bar B^*$ thresholds, and they
can decay via quark rearrangement. Our calculation suggests that no stable $cb \bar u \bar d$ state does exist.

Together with the mass spectra, the wave functions are also obtained by solving the generalized eigenvalue problem of Hamiltonian. With
these wave functions, we can calculate the proportions of hidden color components and the root mean square radii. Besides the
$|\bar 3 3\rangle$ and $|6 \bar 6\rangle$ classifications, one can also define other sets of color representations,
\begin{equation}
|11\rangle = |(Q_1 \bar q_3)^1 (Q_2^\prime \bar q_4^\prime)^1\rangle,
\end{equation}
\begin{equation}
|88\rangle = |(Q_1 \bar q_3)^8 (Q_2^\prime \bar q_4^\prime)^8\rangle,
\end{equation}
and
\begin{equation}
|1^\prime 1^\prime\rangle = |(Q_1 \bar q_4^\prime)^1 (Q_2^\prime \bar q_3)^1\rangle,
\end{equation}
\begin{equation}
|8^\prime 8^\prime\rangle = |(Q_1 \bar q_4^\prime)^8 (Q_2^\prime \bar q_3)^8\rangle.
\end{equation}
Then, the three sets of color representations can be related as follows,
\begin{equation}
|11\rangle = \sqrt{\frac{1}{3}}|\bar 3 3\rangle + \sqrt{\frac{2}{3}}|6 \bar 6\rangle,
\end{equation}
\begin{equation}
|88\rangle = - \sqrt{\frac{2}{3}}|\bar 3 3\rangle + \sqrt{\frac{1}{3}}|6 \bar 6\rangle,
\end{equation}
and
\begin{equation}
|1^\prime 1^\prime\rangle = - \sqrt{\frac{1}{3}}|\bar 3 3\rangle + \sqrt{\frac{2}{3}}|6 \bar 6\rangle,
\end{equation}
\begin{equation}
|8^\prime 8^\prime\rangle = \sqrt{\frac{2}{3}}|\bar 3 3\rangle + \sqrt{\frac{1}{3}}|6 \bar 6\rangle.
\end{equation}
Here, we adopt the $|11\rangle$ and $|88\rangle$ representations to stand for the neutral color and hidden color components, respectively.

\begin{figure}[!htbp]
\includegraphics[scale=1.0]{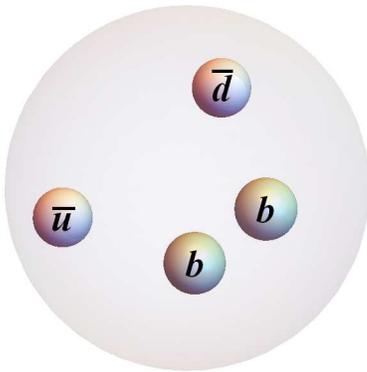}
\vspace{0.0cm} \caption{The stable $IJ^P=01^+$ $bb\bar u \bar d$ state.}
\label{bbud}
\end{figure}

The color proportions and root mean square radii of the three lowest $cc\bar u \bar d$, $bb\bar u \bar d$, and $cb\bar u \bar d$
states are presented in Table~\ref{pro1}. The large hidden color component and small root mean square radius indicate that the
$IJ^P=01^+$ $bb\bar u \bar d$ state is a compact tetraquark rather than a loosely bound molecule. Also, the
$0.285 \sim 0.484~\rm{fm}$ radii differentiate it from a point-like diquark-antidiquark structure. The sketch of this stable
$T_{QQ^\prime}$ state is presented in Figure~\ref{bbud}. It can be seen that the two heavy quarks stay close to each other like a
static color source, while the light antiquark pair circles around this source and is shared by two heavy quarks.

\begin{table*}[htp]
\begin{center}
\caption{\label{pro1} The color proportions and the root mean square radii of the three lowest $cc\bar u \bar d$, $bb\bar u \bar d$,
and $cb\bar u \bar d$ states. The expectations $\langle \boldsymbol r_{14}^2  \rangle^{1/2}$, $\langle \boldsymbol r_{23}^2
\rangle^{1/2}$, and $\langle \boldsymbol r^{\prime \prime 2} \rangle^{1/2}$ equal to the values of $\langle
\boldsymbol r_{24}^2  \rangle^{1/2}$, $\langle \boldsymbol r_{13}^2  \rangle^{1/2}$, and $\langle \boldsymbol r^{\prime 2} \rangle^{1/2}$,
respectively, which are omitted for simplicity. The units of masses and root mean square radii are in MeV and fm, respectively.}
\begin{tabular*}{18cm}{@{\extracolsep{\fill}}*{12}{p{1.3cm}<{\centering}}}
\hline\hline
 System  & Mass  &   $|\bar 3 3\rangle$  &  $|6 \bar 6\rangle$  &  $|11\rangle$  & $|88\rangle$  &  $\langle \boldsymbol r_{12}^2
 \rangle^{1/2}$    &  $\langle \boldsymbol r_{34}^2 \rangle^{1/2}$   &  $\langle \boldsymbol r^2 \rangle^{1/2}$  &
 $\langle \boldsymbol r_{13}^2  \rangle^{1/2}$   &  $\langle \boldsymbol r_{24}^2  \rangle^{1/2}$
 &  $\langle \boldsymbol r^{\prime 2} \rangle^{1/2}$  \\\hline
 $\{c c\}[\bar u \bar d]$  & 4041  &   95.8\%  &  4.2\%  &  34.7\%  & 65.3\%   &  0.449  &  0.597     &  0.386  &  0.537  &  0.537
 &  0.402 \\
 $\{bb\}[\bar u \bar d]$  & 10550  &   99.8\%  &  0.2\%  &  33.4\%  & 66.6\%   &  0.285  &  0.484     &  0.370  &  0.465  &  0.465
 &  0.274 \\
$(cb)[\bar u \bar d]$  & 7297  &   94.0\%  &  6.0\%  &  35.3\%  & 64.7\%       &  0.357  &  0.489     &  0.373  &  0.521  &  0.455
 &  0.324 \\
\hline\hline
\end{tabular*}
\end{center}
\end{table*}

\subsection{Strange systems}

\begin{figure*}[!htbp]
\includegraphics[scale=0.5]{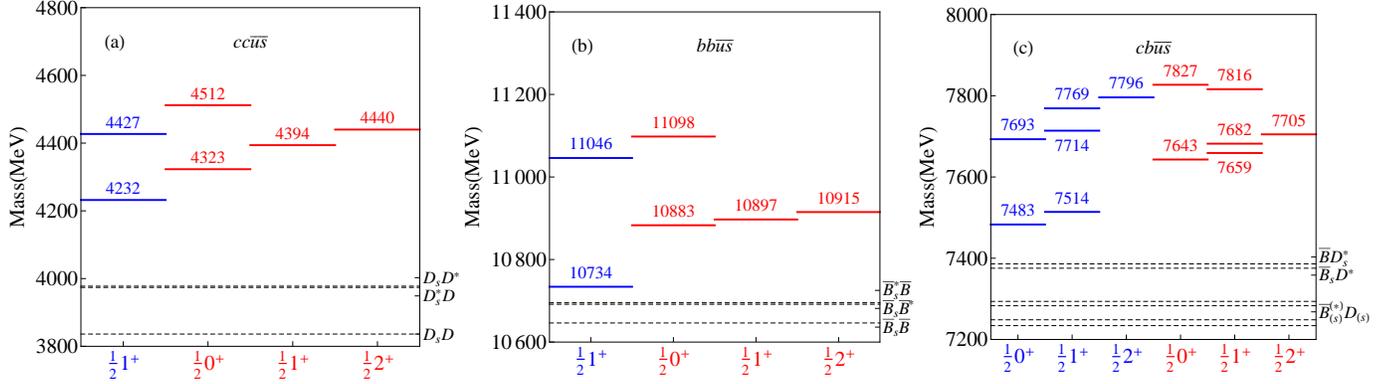}
\vspace{-0.5cm} \caption{The predicted masses of $cc\bar u \bar s$, $bb\bar u \bar s$, and $cb\bar u \bar s$ systems together with
relevant thresholds. The blue lines stand for the tetraquarks including antisymmetric light subsystem $[\bar u \bar s]$, and the red
lines correspond to the ones with symmetric light subsystem $\{\bar u \bar s\}$.}
\label{us}
\end{figure*}

In present work, we treat the antisymmetric $[\bar u \bar s]$ and symmetric $\{\bar u \bar s\}$ as different flavor parts and
do not consider the admixture between them. This situation is similar as the conventional $\Xi_{c(b)}$ and $\Xi_{c(b)}^\prime$ baryons,
which are usually regarded as two independent families. The mass spectra for the  $cc\bar u \bar s$, $bb\bar u \bar s$, and
$cb\bar u \bar s$ systems are shown in Table~\ref{mass2} and Figure~\ref{us}. All of the tetraquarks locate above the
corresponding thresholds, and the three lowest ones for these systems are 4232, 10734, and 7483 MeV, respectively. Analogously,
the $0^+$ and $2^+$ states can decay into a pair of pseudoscalar mesons, and the $1^+$ states can fall apart into a vector meson
plus a pseudoscalar one.

\begin{table*}[htp]
\begin{center}
\caption{\label{mass2} Predicted mass spectra for the $cc\bar u \bar s$, $bb\bar u \bar s$, and $cb\bar u \bar s$ systems.}
\begin{tabular*}{18cm}{@{\extracolsep{\fill}}*{5}{p{3.3cm}<{\centering}}}
\hline\hline
 $IJ^P$  & Configuration                                             & $\langle H\rangle$ (MeV) & Mass (MeV)  & Eigenvector\\\hline
 $\frac{1}{2}1^+$   & $|\{cc\}^{\bar 3}_1 [\bar u \bar s]^3_0\rangle_1$       & \multirow{2}{*}{$\begin{pmatrix}4246&-50
 \\-50&4414\end{pmatrix}$}
               & \multirow{2}{*}{$\begin{bmatrix}4232 \\4427 \end{bmatrix}$}  & \multirow{2}{*}{$\begin{bmatrix}(-0.965, -0.263)\\(0.263,
               -0.965)\end{bmatrix}$}\\
                & $|\{cc\}^6_0 [\bar u \bar s]^{\bar 6}_1\rangle_1$ \\
 $\frac{1}{2}0^+$  &  $|\{cc\}^{\bar 3}_1 \{\bar u \bar s\}^3_1\rangle_0$     & \multirow{2}{*}{$\begin{pmatrix}4370&-82
 \\-82&4465\end{pmatrix}$}
               & \multirow{2}{*}{$\begin{bmatrix}4323 \\4512 \end{bmatrix}$}  & \multirow{2}{*}{$\begin{bmatrix}(-0.865, -0.501)\\(0.501,
               -0.865)\end{bmatrix}$}\\
                 &  $|\{cc\}^6_0 \{\bar u \bar s\}^{\bar 6}_0\rangle_0$    \\
 $\frac{1}{2}1^+$  &  $|\{cc\}^{\bar 3}_1 \{\bar u \bar s\}^3_1\rangle_1$    & 4394  &  4394  &  1\\
 $\frac{1}{2}2^+$  &  $|\{cc\}^{\bar 3}_1 \{\bar u \bar s\}^3_1\rangle_2$    & 4440  &  4440  &  1\\\hline

 $\frac{1}{2}1^+$   & $|\{bb\}^{\bar 3}_1 [\bar u \bar s]^3_0\rangle_1$       & \multirow{2}{*}{$\begin{pmatrix}10736&-19
 \\-19&11044\end{pmatrix}$}
               & \multirow{2}{*}{$\begin{bmatrix}10734 \\11046 \end{bmatrix}$}  & \multirow{2}{*}{$\begin{bmatrix}(-0.998, -0.060)\\
               (0.060, -0.998)\end{bmatrix}$}\\
                & $|\{bb\}^6_0 [\bar u \bar s]^{\bar 6}_1\rangle_1$ \\
 $\frac{1}{2}0^+$  &  $|\{bb\}^{\bar 3}_1 \{\bar u \bar s\}^3_1\rangle_0$     & \multirow{2}{*}{$\begin{pmatrix}10888
 &29 \\29&11094\end{pmatrix}$}
               & \multirow{2}{*}{$\begin{bmatrix}10883 \\11098 \end{bmatrix}$}  & \multirow{2}{*}{$\begin{bmatrix}(-0.990, 0.138)\\
               (-0.138, -0.990)\end{bmatrix}$}\\
                 &  $|\{bb\}^6_0 \{\bar u \bar s\}^{\bar 6}_0\rangle_0$    \\
 $\frac{1}{2}1^+$  &  $|\{bb\}^{\bar 3}_1 \{\bar u \bar s\}^3_1\rangle_1$    & 10897  &  10897  &  1\\
 $\frac{1}{2}2^+$  &  $|\{bb\}^{\bar 3}_1 \{\bar u \bar s\}^3_1\rangle_2$     & 10915  &  10915  &  1\\\hline

 $\frac{1}{2}0^+$  &  $|(cb)^{\bar 3}_0 [\bar u \bar s]^3_0 \rangle_0$      & \multirow{2}{*}{$\begin{pmatrix}7502
 &61 \\61&7673\end{pmatrix}$}
               & \multirow{2}{*}{$\begin{bmatrix}7483 \\7693 \end{bmatrix}$}  & \multirow{2}{*}{$\begin{bmatrix}(-0.952, 0.306)\\
               (-0.306, -0.952)\end{bmatrix}$}\\
               &  $|(cb)^6_1 [\bar u \bar s]^{\bar 6}_1\rangle_0$ \\
 $\frac{1}{2}1^+$ &  $|(cb)^{\bar 3}_1 [\bar u \bar s]^3_0\rangle_1$      & \multirow{3}{*}{$\begin{pmatrix}7519&32&15 \\32&7761
 &-16 \\15& -16 & 7717\end{pmatrix}$}
               & \multirow{3}{*}{$\begin{bmatrix}7514 \\7714 \\ 7769 \end{bmatrix}$}  & \multirow{3}{*}{$\begin{bmatrix}
               (-0.987, 0.134, 0.086)\\(0.117, 0.249, 0.961) \\ (0.107, 0.959, -0.262)\end{bmatrix}$}\\
                &  $|(cb)^6_0 [\bar u \bar s]^{\bar 6}_1\rangle_1$     \\
              &  $|(cb)^6_1 [\bar u \bar s]^{\bar 6}_1\rangle_1$ \\
$\frac{1}{2}2^+$ &  $|(cb)^6_1 [\bar u \bar s]^{\bar 6}_1\rangle_2$    & 7796  &  7796  &  1\\

 $\frac{1}{2}0^+$  & $|(cb)^{\bar 3}_1 \{\bar u \bar s\}^3_1\rangle_0$    & \multirow{2}{*}{$\begin{pmatrix}7659
 &52 \\52&7811\end{pmatrix}$}
               & \multirow{2}{*}{$\begin{bmatrix}7643 \\7827 \end{bmatrix}$}  & \multirow{2}{*}{$\begin{bmatrix}(-0.955, 0.297)\\
               (-0.297, -0.955)\end{bmatrix}$}\\
                 &  $|(cb)^6_0 \{\bar u \bar s\}^{\bar 6}_0\rangle_0$    \\
 $\frac{1}{2}1^+$  &  $|(cb)^{\bar 3}_0 \{\bar u \bar s\}^3_1\rangle_1$     & \multirow{3}{*}{$\begin{pmatrix}7674&-9&-30 \\
 -9&7675&-14 \\ -30& -14 & 7808\end{pmatrix}$}
               & \multirow{3}{*}{$\begin{bmatrix}7659 \\7682 \\ 7816 \end{bmatrix}$}  & \multirow{3}{*}{$\begin{bmatrix}
               (0.769, 0.604, 0.211)\\(0.608, -0.792, 0.053) \\ (-0.199, -0.087, 0.976)\end{bmatrix}$}\\
                &  $|(cb)^{\bar 3}_1 \{\bar u \bar s\}^3_1\rangle_1$     \\
              &    $|(cb)^6_1 \{\bar u \bar s\}^{\bar 6}_0\rangle_1$         \\
$\frac{1}{2}2^+$    &  $|(cb)^{\bar 3}_1 \{\bar u \bar s\}^3_1\rangle_2$    & 7705  &  7705  &  1\\

\hline\hline
\end{tabular*}
\end{center}
\end{table*}

It should be mentioned that in the literature some results supported a stable $\{bb\}[\bar u \bar s]$ state with
$IJ^P=\frac{1}{2}1^+$~\cite{Deng:2018kly,Eichten:2017ffp,Francis:2016hui,SilvestreBrac:1993ss,Du:2012wp,Junnarkar:2018twb}, and others
predicted a state near the open bottom thresholds~\cite{Park:2018wjk,Ebert:2007rn}. Our results show that the lowest
$\{bb\}[\bar u \bar s]$ state is about 40 MeV above the $\bar B_s \bar B^*$ and $\bar B^*_s \bar B$ thresholds. Considering the uncertainties of relativized quark model, we conclude that a resonance-like $\{bb\}[\bar u \bar s]$ structure
may exist. The results of color proportions and root mean square radii of the three lowest $cc\bar u \bar s$, $bb\bar u \bar s$, and
$cb\bar u \bar s$ states are also listed in Table~\ref{pro2} for reference. More experimental searches are expected to resolve this
problem in the future.

\begin{table*}[htp]
\begin{center}
\caption{\label{pro2} The color proportions and root mean square radii of the three lowest $cc\bar u \bar s$, $bb\bar u \bar s$,
and $cb\bar u \bar s$ states. The units of masses and root mean square radii are in MeV and fm, respectively.}
\begin{tabular*}{18cm}{@{\extracolsep{\fill}}*{12}{p{1.3cm}<{\centering}}}
\hline\hline
 System  & Mass  &   $|\bar 3 3\rangle$  &  $|6 \bar 6\rangle$  &  $|11\rangle$  & $|88\rangle$  &  $\langle \boldsymbol r_{12}^2
 \rangle^{1/2}$    &  $\langle \boldsymbol r_{34}^2 \rangle^{1/2}$   &  $\langle \boldsymbol r^2 \rangle^{1/2}$  &
 $\langle \boldsymbol r_{13}^2  \rangle^{1/2}$   &  $\langle \boldsymbol r_{24}^2  \rangle^{1/2}$   &  $\langle \boldsymbol r^{\prime 2}
 \rangle^{1/2}$  \\\hline
$\{cc\}[\bar u \bar s]$  & 4232  &   93.1\%  &  6.9\%  &  35.6\%  & 64.4\%  &  0.423  &  0.491     & 0.384  &  0.544  &  0.470  &  0.363 \\
$\{bb\}[\bar u \bar s]$  & 10734  &   99.6\%  &  0.4\%  &  33.5\%  & 66.5\%  &  0.284  &  0.484      &  0.364  &  0.503  &  0.425
&  0.269 \\
$(cb)[\bar u \bar s]$  & 7483  &   90.6\%  &  9.4\%  &  36.5\%  & 63.5\%  &  0.358  &  0.493       &  0.365  &  0.557  &  0.412
 &  0.324 \\
\hline\hline
\end{tabular*}
\end{center}
\end{table*}

\begin{figure*}[!htbp]
\includegraphics[scale=0.5]{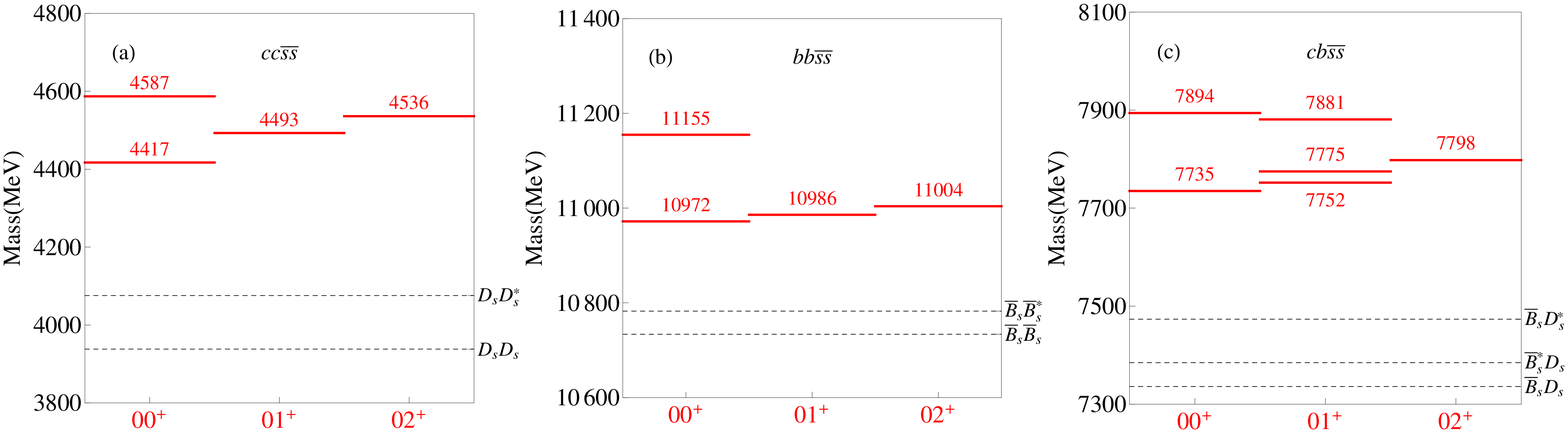}
\vspace{-0.5cm} \caption{The predicted masses of the $cc\bar s \bar s$, $bb\bar s \bar s$, and $cb\bar s \bar s$ systems together
with relevant thresholds.}
\label{ss}
\end{figure*}

For the $cc\bar s \bar s$, $bb\bar s \bar s$, and $cb\bar s \bar s$ systems, the strange quark pair must be symmetric in flavor part
and therefore, less states are predicted. From Table~\ref{mass3} and Figure~\ref{ss}, It can been seen that all of them lie much
higher than the corresponding thresholds and can easily fall apart into the charmed strange or bottom strange final states. Our
results are consistent with other theoretical works~\cite{Ebert:2007rn,Zhang:2007mu}, and we believe that no stable structure exists
in $cc\bar s \bar s$, $bb\bar s \bar s$, and $cb\bar s \bar s$ systems.

\begin{table*}[htp]
\begin{center}
\caption{\label{mass3} Predicted mass spectra for the $cc\bar s \bar s$, $bb\bar s \bar s$, and $cb\bar s \bar s$ systems.}
\begin{tabular*}{18cm}{@{\extracolsep{\fill}}*{5}{p{3.3cm}<{\centering}}}
\hline\hline
 $IJ^P$  & Configuration                                             & $\langle H\rangle$ (MeV) & Mass (MeV)  & Eigenvector\\\hline
 $00^+$  &  $|\{cc\}^{\bar 3}_1 \{\bar s \bar s\}^3_1\rangle_0$     & \multirow{2}{*}{$\begin{pmatrix}4469&79 \\79&4535\end{pmatrix}$}
               & \multirow{2}{*}{$\begin{bmatrix}4417 \\4587 \end{bmatrix}$}  & \multirow{2}{*}{$\begin{bmatrix}(-0.832, 0.555)\\
               (-0.555, -0.832)\end{bmatrix}$}\\
                 &  $|\{cc\}^6_0 \{\bar s \bar s\}^{\bar 6}_0\rangle_0$    \\
 $01^+$  &  $|\{cc\}^{\bar 3}_1 \{\bar s \bar s\}^3_1\rangle_1$    & 4493  &  4493  &  1\\
 $02^+$  &  $|\{cc\}^{\bar 3}_1 \{\bar s \bar s\}^3_1\rangle_2$    & 4536  &  4536  &  1\\\hline

 $00^+$  &  $|\{bb\}^{\bar 3}_1 \{\bar s \bar s\}^3_1\rangle_0$     & \multirow{2}{*}{$\begin{pmatrix}10977&-29 \\-29&11151\end{pmatrix}$}
               & \multirow{2}{*}{$\begin{bmatrix}10972 \\11155 \end{bmatrix}$}  & \multirow{2}{*}{$\begin{bmatrix}(-0.987, -0.159)\\
               (0.159, -0.987)\end{bmatrix}$}\\
                 &  $|\{bb\}^6_0 \{\bar s \bar s\}^{\bar 6}_0\rangle_0$    \\
 $01^+$  &  $|\{bb\}^{\bar 3}_1 \{\bar s \bar s\}^3_1\rangle_1$    & 10986  &  10986  &  1\\
 $02^+$  &  $|\{bb\}^{\bar 3}_1 \{\bar s \bar s\}^3_1\rangle_2$    & 11004  &  11004  &  1\\\hline

 $00^+$  & $|(cb)^{\bar 3}_1 \{\bar s \bar s\}^3_1\rangle_0$    & \multirow{2}{*}{$\begin{pmatrix}7753&-50 \\-50&7876\end{pmatrix}$}
               & \multirow{2}{*}{$\begin{bmatrix}7735 \\7894 \end{bmatrix}$}  & \multirow{2}{*}{$\begin{bmatrix}(-0.941, -0.337)\\
               (0.337, -0.941)\end{bmatrix}$}\\
                 &  $|(cb)^6_0 \{\bar s \bar s\}^{\bar 6}_0\rangle_0$    \\
 $01^+$  &  $|(cb)^{\bar 3}_0 \{\bar s \bar s\}^3_1\rangle_1$     & \multirow{3}{*}{$\begin{pmatrix}7767&-8&29 \\-8&7769&13 \\ 29& 13
 & 7873\end{pmatrix}$}
               & \multirow{3}{*}{$\begin{bmatrix}7752 \\7775 \\ 7881 \end{bmatrix}$}  & \multirow{3}{*}{$\begin{bmatrix}
               (0.784, 0.570, -0.248)\\(-0.576, 0.816, 0.056) \\ (-0.234, -0.099, -0.967)\end{bmatrix}$}\\
                &  $|(cb)^{\bar 3}_1 \{\bar s \bar s\}^3_1\rangle_1$     \\
              &   $|(cb)^6_1 \{\bar s \bar s\}^{\bar 6}_0\rangle_1$     \\
$02^+$    &  $|(cb)^{\bar 3}_1 \{\bar s \bar s\}^3_1\rangle_2$    & 7798  &  7798  &  1\\

\hline\hline
\end{tabular*}
\end{center}
\end{table*}

\subsection{Mass ratios}

\begin{figure*}[!htbp]
\includegraphics[scale=0.5]{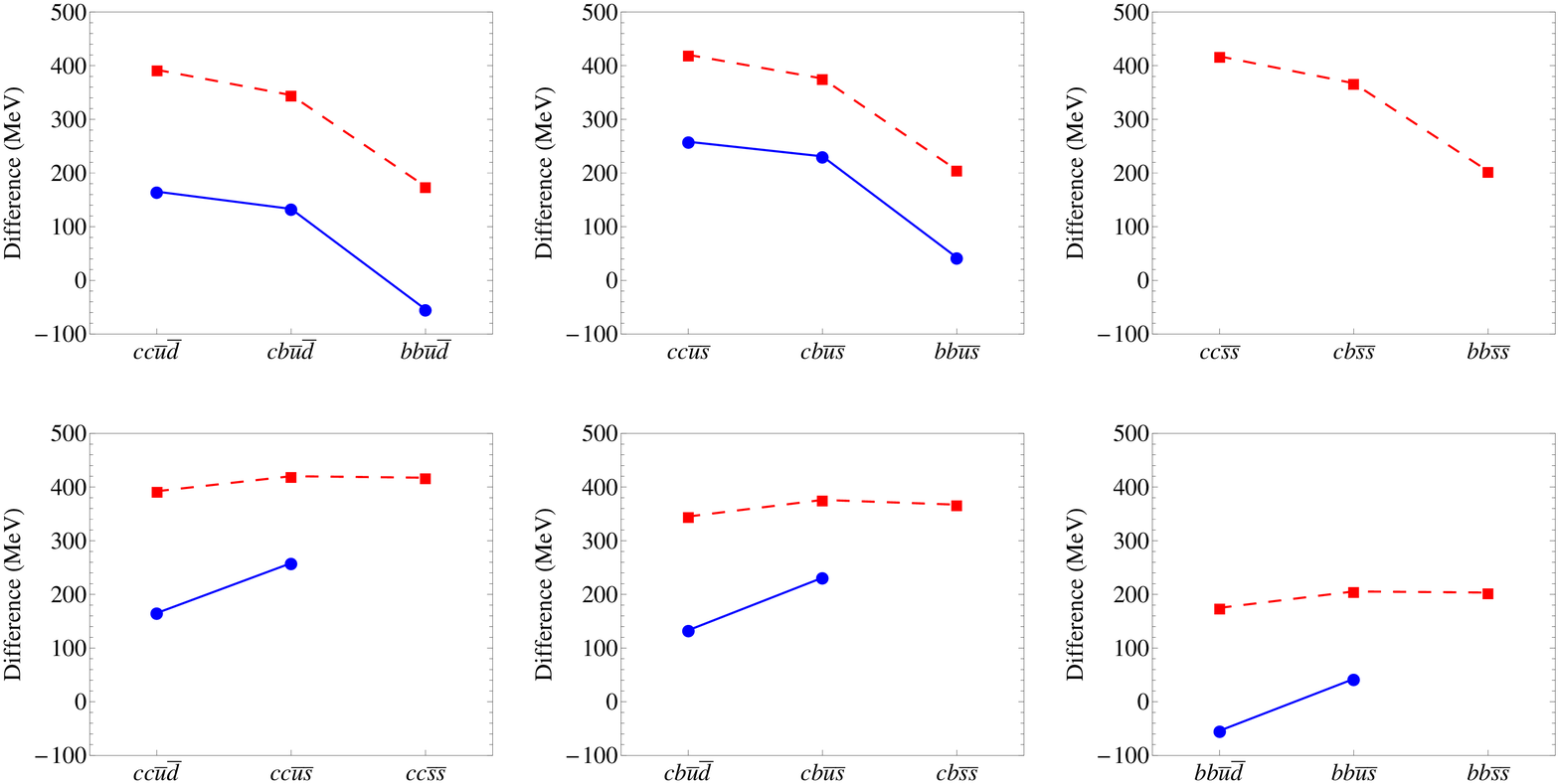}
\vspace{-0.5cm} \caption{Mass differences between lower $J^P=1^+$ tetraquarks and thresholds versus the different systems. The
blue points stand for the tetraquarks including antisymmetric light subsystems, and the red squares correspond to the ones with
symmetric light subsystems.}
\label{difference}
\end{figure*}

With the mass spectra of the doubly heavy tetraquarks $T_{QQ^\prime}$, one can discuss the mass differences between tetraquark states
and the corresponding thresholds. For instance, the mass differences between lower $J^P=1^+$ tetraquarks and thresholds versus the
different systems are plotted in Figure~\ref{difference}. With the fixed light antiquark subsystem, the mass differences decrease
when the heavy quarks vary from $cc$ to $bb$. Similarly, for a certain heavy quark subsystem, the mass differences show upward trends
when the light antiquarks change from the $\bar u \bar d$ to $\bar s \bar s$. The $IJ^P=01^+$ $\{bb\}[\bar u \bar d]$ state has the
largest mass ratio between heavy quarks and light antiquarks, which forms a binding compact tetraquark. With the mass ratios between
two subsystems decreasing, we can not obtain stable doubly heavy tetraquarks.

In Refs.~\cite{Vijande:2009kj,Hernandez:2019eox}, the authors also discussed the dependence of mass ratios between the heavy and
light subsystems within nonrelativistic quark model, and showed the same behaviors with our relativized calculations. If one keeps
reducing the mass ratios, the doubly heavy tetraquarks will become fully heavy tetraquarks. We can speculate that there is no stable
state for the fully heavy tetraquarks since the mass ratios between the two subsystems are sufficiently small. This conjecture is
supported by the experimental observations~\cite{Aaij:2018zrb,Sirunyan:2020txn} and nonrelativistic quark model works with proper
potentials~\cite{Richard:2017vry,Liu:2019zuc,Wang:2019rdo,Chen:2019vrj,Richard:2019cmi,Deng:2020iqw}. Certainly, the classifications
of fully heavy tetraquarks are different with doubly heavy tetraquarks, precise calculations within the relativized quark model are
needed before coming to any conclusion.

\section{Summary}{\label{Summary}}

In this work, we systematically investigate the mass spectra of doubly heavy tetraquarks $T_{QQ^\prime}$ in a relativized quark model.
The four-body systems including the Coulomb potential, confining potential, spin-spin interactions, and relativistic corrections are
solved within the variational method. With the present extension, the tetraquark, as well as the conventional hadrons can be described in a uniform frame. Our results suggest that the $IJ^P=01^+$ $bb \bar u \bar d$ state
is 54 MeV below the relevant $\bar B \bar B$ and $\bar B \bar B^*$ thresholds, which indicates that both strong and electromagnetic decays are forbidden, and thus this state can be a stable one. The large hidden
color component and small root mean square radius demonstrate that it is a compact tetraquark rather than a loosely bound molecule or
point-like diquark-antidiquark structure. Compared with the results of nonrelativistic quark models, our calculations present
a lower binding energy of this promising isoscalar $T_{bb}$ state, but the decay behaviors agree with each other. We believe
our calculations and  predictions of the doubly heavy tetraquarks may provide valuable information for future experimental searches.

\bigskip
\noindent
\begin{center}
{\bf ACKNOWLEDGEMENTS}\\
\end{center}
We would like to thank Xian-Hui Zhong, Ming-Sheng Liu, and Wei Liang for helpful discussions. This project is supported by the National Natural Science Foundation of China under Grants No.~11705056, No.~11775050, No.~11947224, No.~11975245, and No.~U1832173, by the fund provided to the Sino-German CRC 110
"Symmetries and the Emergence of Structure in QCD" project by the NSFC under Grant No.~11621131001,
and by the Key Research Program of Frontier Sciences, CAS, Grant No. Y7292610K1.

\end{document}